# An Improved SIW Leaky Wave Antenna Based on a Compact CRLH Unit Cell

Hossein Dolatkhah [1], Zahra Atlasbaf [1*]

[1] Department of Electrical and Computer Engineering, Tarbiat Modares University, Tehran, 14115-111, Iran
[*] atlasbaf@modares.ac.ir

**Abstract:** In this paper, a new composite right/left-handed (CRLH) Leaky Wave Antenna (LWA) based on substrate-integrated waveguide (SIW) structures is proposed. The suggested antenna has a continuous beam scanning from backward to forward versus frequency and can be used for wideband applications. The compact composite right/left-handed (C-CRLH) unit cell for the mentioned antenna is designed by etching two transverse slots and one interdigital slot between them on the upper ground plate of the SIW structure. The C-CRLH unit cell dispersion diagram extracted from the simulation results is compared with the transmission line approach results. The CRLH band is from 12GHz to 24GHz, and the antenna operation frequency band is from 13GHz to 22GHz. The beam scanning angle of the proposed antenna is from $-53^0$ to $+66^0$. The maximum radiation efficiency is 91.89% and more than 85.7% in 13GHz-19.5GHz. It is worth noting that the maximum gain of the antenna is 12.66dBi, and it is approximately 12 dBi in 13GHz-20GHz. The measurement results of the antenna are compared with simulation results, and a good agreement is observed.

## 1. Introduction

Recently, leaky-wave antennas (LWAs) have become interesting structures. Their beam scanning capability against frequency makes them an attractive choice for microwave applications, especially scanning applications. In common LWAs, beam scanning is only possible in the forward direction [1], but in modern applications, we need LWAs capable of continuous scanning from the backward to forward direction [2]. Consequently, composite right-left-handed (CRLH) metamaterials are employed in the LWAs structure to acquire continuous beam scanning from backward to forward. The backward direction scanning property of metamaterials is a result of their negative permittivity and permeability in the left-hand region [2].

The substrate integrated waveguide (SIW) is also an essential component of most LWAs because it has a low profile, can be integrated into planar circuits, and is a low-cost structure [3], [4], [5]. Further, metallic vias in SIW play a vital role in the creation of backward beam steering conditions. In conclusion, all of these lead to the use of SIW structure [2], [6].

Recently, LWAs with different types of Single-Periodic CRLH (SP-CRLH) and Double-Periodic CRLH (DP-CRLH) unit cells based on SIW technology have been studied in [6]-[11]. In the referenced works, various CRLH unit-cells are developed to enable backward to forward beam steering capability in the conventional LWAs and to promote other characteristics like gain, radiation efficiency, and bandwidth. In [6], CRLH-SIW and half mode SIW (HMSIW) leaky-wave structures for antenna applications are proposed. The used unit cells in these structures are realized by engendering interdigital slots on the surface of the SIW structure for one-side radiation and on the surface and ground for two-side radiation. The designed antennas can steer the beam from backward to forward and the final developed antenna scans the beam from $-70^0$ to $+60^0$. A multilayered LWA using SIW technology is proposed in [7] to achieve the continuous beam scanning from $-66^0$ to $+78^0$ with consistent gain. The CRLH unit cell is formed by cutting a transverse slot on the upper ground of SIW, and a parasitic patch under the transverse slot. A two-layered wideband LWA is developed in [8] for radar systems and imaging applications. The proposed CRLH unit cell for this structure consists of an interdigital slot and a pair of twisted inductive posts. The impedance bandwidth of the antenna is from 6.5 GHz to 9.5 GHz with frequency beam scanning from $-34^0$ to $+72^0$. In [9], a circularly polarized SIW LWA loaded by longitudinal and transverse slots is presented with beam steering from $-40^0$ to $+35^0$ and the maximum gain of 12 dBi. A CRLH LWA based on SIW technology with low cross-polarization and the continuous beam scanning from backward to forward is developed and presented in [10]. The proposed antenna has frequency beam scanning from $-74^0$ to $+45^0$ with the measured cross-polarization level smaller than -20.8 dB which is acquired by using the combined cross-shaped and transverse slots in the developed CRLH unit cell structure. In [11], two adjacent interdigital slots with two vias between them are used to develop a new DP-CRLH and design SIW based LWA. The proposed antenna based on the new DP-CRLH has two CRLH bands and a new right-handed (RH) band. In the first CRLH band beam scanning is from $-78^0$ to $+78^0$ with antenna radiation efficiency of about 90%. Beam steering in the second CRLH band which is an unbalanced band is from $-40^0$ to $+20^0$ and in the RH band is from $+22^0$ to $+54^0$. Furthermore, [12] presents a SIW LW slot antenna with a modified single periodic CRLH unit cell for improving scanning range and gain flatness. A beam scanning range of $-17^0$ to $+13^0$ is achieved by using the modified SP-CRLH unit cell, which consists of two longitude slots and a meandered slot between them.

In this paper, we present a SIW based LWA with a developed compact CRLH (C-CRLH) unit cell. The C-CRLH unit cell consists of transverse and interdigital slots. The transverse and interdigital loaded unit cell in the present work has a wide impedance matching range. Besides, it provides a wide beam steering range. Furthermore, the developed antenna has better performance than the reference antennas in most of the features that are listed in Table 2, because of the developed C-CRLH unit cell. It is noteworthy that in this design, the combination of the interdigital and transverse slots is used as a novel CRLH unit cell. However, each of these slots is presented as a novel CRLH unit cell in some of the reported literature as well as in [13-30].

This paper is prepared as follows: In the first section, the SIW LWA based on the developed C-CRLH unit cell is described. The unit cell circuit model and dispersion analysis are also discussed in this section. In the second Section, the measured results of the fabricated antenna are compared to the simulated results and discussed in detail. Finally, the last section concludes the paper.

## 2. Developed Structure

### 2.1. Compact CRLH (C-CRLH) unit cell

The proposed C-CRLH unit is shown in Fig. 1. It consists of an interdigital and two transverse slots. A SIW structure loaded by either an interdigital slot or a transverse slot can be a CRLH unit cell; hence, the proposed structure in Fig. 1, behaves like a three-in-one unit cell with a periodicity of T, where T is the period of an SP-CRLH. The CRLH unit cell is designed on Rogers RT 5880 with a permittivity of 2.2, a loss tangent of 0.0009, and thickness h=0.8mm.

The equivalent circuit model of the developed unit cell is depicted in Fig. 2. In Fig. 2(a), the SP-CRLH unit cell's circuit model is shown. The SIW structure is a transmission line with distributed series inductance (L_R1) and shunt capacitance (C_R1). The left-handed (LH) series capacitance (C_L1) can be provided by cutting either transverse or interdigital slot on the surface of the SIW, and the shunt inductance (L_L1) is provided by the SIW metallic vias [2]. The circuit model of the C-CRLH unit cell is pictured in Fig. 2(b), which consists of three SP-CRLH circuit models.

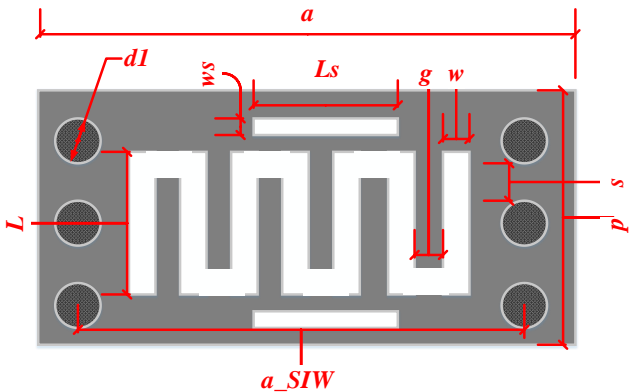

**Fig. 1.** *The Developed C-CRLH unit cell.*

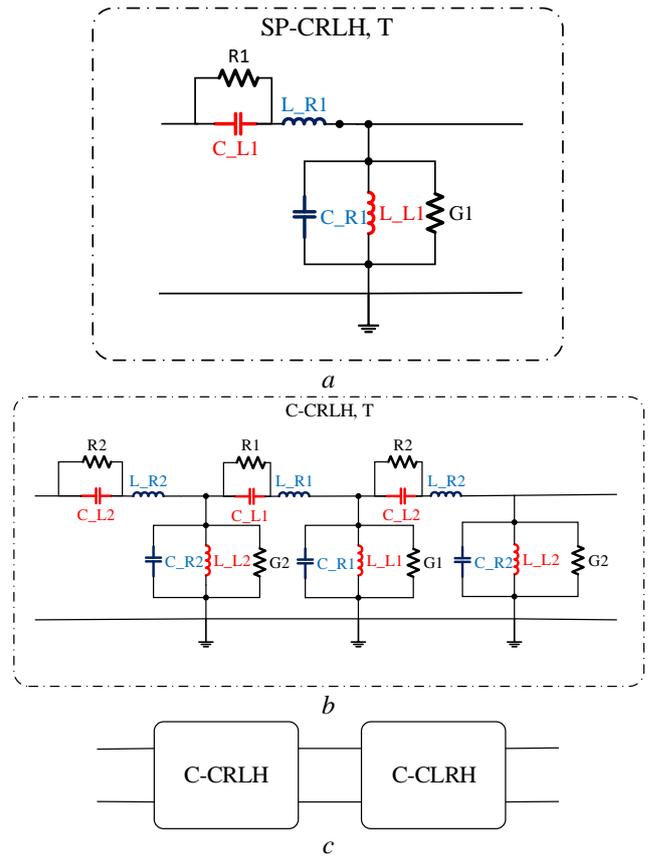

**Fig. 2.** *Equivalent circuit model of the structure. (a)* SP-CRLH unit cell, **(b)** C-CRLH unit cell, **(c)** Line implementation in form of the periodic network.

### 2.2 Transmission line approach and dispersion analysis

The transmission line approach is the method that can be used for transmission matrix extraction in the CRLH based unit cells [2].

The transmission matrix of Fig. 2(b) can be derived by multiplying the transmission matrixes of the three different SP-CRLH unit cells [21] as

$$\begin{bmatrix} V_n \\ I_n \end{bmatrix} = U_1 * U_2 * U_3 * \begin{bmatrix} V_{n+1} \\ I_{n+1} \end{bmatrix} = \begin{bmatrix} A & B \\ C & D \end{bmatrix} \begin{bmatrix} V_{n+1} \\ I_{n+1} \end{bmatrix} \quad (1)$$

where

$$U_1 = U_3 = \begin{bmatrix} 1 & Z_{R2} \| Z_{C\_L2} \\ 0 & 1 \end{bmatrix} \begin{bmatrix} 1 & Z_{L\_R2} \\ 0 & 1 \end{bmatrix} \begin{bmatrix} 1 & 0 \\ Y_{C\_R2} & 1 \end{bmatrix} \begin{bmatrix} 1 & 0 \\ Y_{L\_L2} & 1 \end{bmatrix} \begin{bmatrix} 1 & 0 \\ Y_{G2} & 1 \end{bmatrix} \quad (2)$$

and

$$U_2 = \begin{bmatrix} 1 & Z_{R_1} \| Z_{C\_L1} \\ 0 & 1 \end{bmatrix} \begin{bmatrix} 1 & Z_{L\_R1} \\ 0 & 1 \end{bmatrix} \begin{bmatrix} 1 & 0 \\ Y_{C\_R1} & 1 \end{bmatrix} \begin{bmatrix} 1 & 0 \\ Y_{L\_L1} & 1 \end{bmatrix} \begin{bmatrix} 1 & 0 \\ Y_{G1} & 1 \end{bmatrix} \quad (3)$$

$U_i$ (i=1, 2, 3) is the ABCD matrix of the SP-CRLH unit cell which is extracted from [13]. $Z_{C\_Li} = 1/j\omega C\_Li$ and $Y_{L\_Li} = 1/j\omega L\_Li$ are the impedance and admittance of the series capacitance and shunt inductance, respectively.

Whereas $Z_{L\_Ri} = j\omega L\_Ri$ and $Y_{C\_Ri} = j\omega C\_Ri$ are the impedance and admittance of the distributed series inductance and shunt capacitance, respectively. Each of the slots can be modeled as radiation resistance ($Z_{Ri} = Ri$). In addition, the antenna conductivity and dielectric losses are modeled as $Y_{Gi} = Gi$. The coefficients A and D are extracted as (4) and (5).

$$A = (Y_{C\_R2} + Y_{Gi} + Y_{L\_L2})(\frac{Z_{C\_L2}Z_{R2}}{Z_{C\_L2} + Z_{R2}} + Z_{L\_R2}) +$$

$$(1 + (Y_{C\_R2} + Y_{G2} + Y_{L\_L2})(\frac{Z_{C\_L2}Z_{R2}}{Z_{C\_L2} + Z_{R2}} + Z_{L\_R2}))$$

$$((Y_{C\_R2} + Y_{G2} + Y_{L\_L2})(\frac{Z_{C\_L1}Z_{R1}}{Z_{C\_L1} + Z_{R1}} + Z_{L\_R1}) +$$

$$(Y_{C\_R1} + Y_{G1} + Y_{L\_L1})(\frac{Z_{C\_L2}Z_{R2}}{Z_{C\_L2} + Z_{R2}} + Z_{L\_R2}) + \quad (4)$$

$$(1 + (Y_{C\_R1} + Y_{G1} + Y_{L\_L1})(\frac{Z_{C\_L1}Z_{R1}}{Z_{C\_L1} + Z_{R1}} + Z_{L\_R1}))$$

$$(1 + (Y_{C\_R2} + Y_{G2} + Y_{L\_L2})(\frac{Z_{C\_L2}Z_{R2}}{Z_{C\_L2} + Z_{R2}} + Z_{L\_R2})))$$

$$D = 1 + (Y_{C\_R2} + Y_{G2} + Y_{L\_L2})(\frac{Z_{C\_L1}Z_{R1}}{Z_{C\_L1} + Z_{R1}} + Z_{L\_R1}) +$$

$$(Y_{C\_R1} + Y_{G1} + Y_{L\_L1})(\frac{Z_{C\_L2}Z_{R2}}{Z_{C\_L2} + Z_{R2}} + Z_{L\_R2}) +$$

$$(Y_{C\_R2} + Y_{G2} + Y_{L\_L2})(1 + (Y_{C\_R1} + Y_{G1} + Y_{L\_L1})$$

$$(\frac{Z_{C\_L1}Z_{R1}}{Z_{C\_L1} + Z_{R1}} + Z_{L\_R1}))(\frac{Z_{C\_L2}Z_{R2}}{Z_{C\_L2} + Z_{R2}} + Z_{L\_R2})$$

(5)

The dispersion relation of the proposed unit cell is calculated based on the bloch-floquet theory [13], [21] and can be written as equation (6)

$$\cos(\beta T) = \frac{A + D}{2} \quad (6)$$

And the block impedance of the unit cell can be calculated as (7) [25]

$$Z_B = \frac{j2Z_0 S_{21} \sin(\beta p)}{(1 - S_{11})(1 - S_{22}) - S_{12}S_{21}} \quad (7)$$

where $Z_0$ (50Ω) is the characteristic impedance and $S_{ij}$ (i=1,2 and j=1,2) is the simulated S-parameter data of the unit cell.

The dimensions of the unit cell are given in Table 1.
Another issue with the unit cell design is the distance between the interdigital slot and the transverse slots (D). By using this parameter, the overall length of the antenna can be calculated. Besides, the optimum value for this parameter can be used to maximize the total radiation efficiency of the antenna [13], [26]. The simulations demonstrated that in the single C-

**Table 1** Dimensions of the designed antenna

| Symbol | Quantity | Value(mm) |
|---|---|---|
| $a$ | Antenna's width | 8 |
| $a\_SIW$ | Antenna's effective width | 7.5 |
| $L$ | Interdigital slots' length | 2 |
| $w$ | Interdigital slots' width | 0.38 |
| $g$ | Interdigital slots' finger width | 0.37 |
| $S$ | Space between adjacent vias | 0.8 |
| $d1$ | Diameter of via | 0.8 |
| $p$ | Period of unit cell | 3.9 |
| $w\_s$ | Transverse slots' width | 0.47 |
| $L\_s$ | Transverse slots' length | 2 |
| $W1$ | 50Ω microstrip line width | 2.5 |
| $W2$ | Tapered line width | 5 |
| $L1$ | 50Ω microstrip line length | 3.7 |
| $L2$ | Tapered line length | 3.2 |

CRLH unit cell, increasing D up to 0.715 mm (p=3.9 mm) increases the coupled power from the C-CLRH unit cell to the free space and vice versa. Meanwhile, increasing D must be done in such a way that p <λ/4 [2]. The coupled power from a unit cell to the free space is defined as (8) [14], [22].

$$CP = 1 - |S_{11}|^2 - |S_{21}|^2 \quad (8)$$

where $|S_{11}|^2$ and $|S_{21}|^2$ are the returned power to the feeding port due to impedance mismatching between the unit cell and the feeding structure, and the absorbed power in the other end side of the unit cell structure, respectively. In this study, numerous simulations have been done to extract S-parameters for the different D and plot the power coupling coefficient. The power coupling coefficient from the C-CRLH unit cell to the free space versus D is plotted in Fig. 3.

Figs. 4 and 5, show the effect of mutual coupling between the transverse straight slots and the interdigital slot on the dispersion diagram and the block impedance, respectively. Based on Fig. 4, as D increases, the balance point of the dispersion diagram moves towards higher frequencies, while when D decreases, the balance point moves towards lower frequencies. Also, the LH region shows some new resonances when D is lower than 0.715mm. These resonances can be caused by decreasing D, which can increase the mutual coupling between slots. According to Fig. 5, there is only a negligible frequency shift in the block impedance diagram, and tuning D does not significantly change either the real or imaginary parts of the block impedance. This work uses 0.715mm as the D value which is an optimum value based on Figs. 3 and 4.

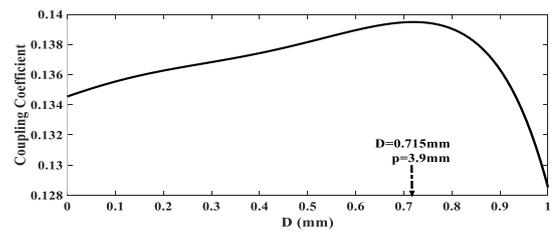

*Fig. 3.  C-CRLH unit cell power coupling to the free space against the interdigital slot and transverse slots distance.*

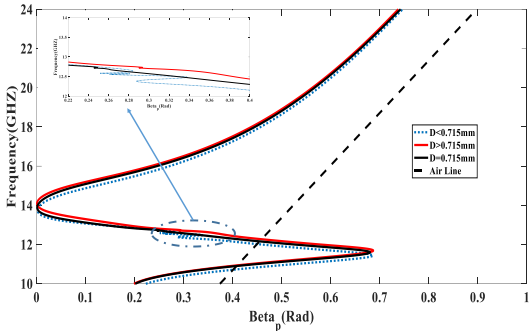

*Fig. 4.* *The effect of mutual coupling between the interdigital slot and transverse slots on the dispersion diagram.*

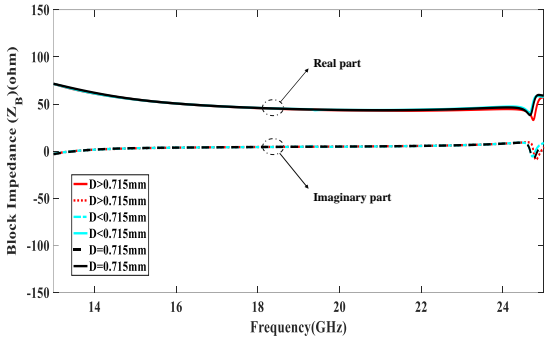

*Fig. 5.* *The effect of mutual coupling between the interdigital slot and transverse slots on the block impedance.*

The extracted dispersion diagram of the C-CRLH unit cell (D=0.715mm) from the transmission line approach and simulation are compared in Fig. 6. As pictured, the circuit model results and the results extracted from the simulation are in good agreement, which confirms that the C-CRLH unit cell behaves similarly to the three SP-CRLH unit cells in series. There is a negligible difference between the two diagrams which can be eliminated by tuning the circuit model parameters and choosing the optimum ones.

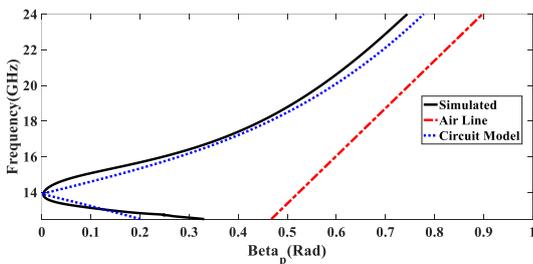

*Fig. 6.* *Dispersion diagram of the developed C-CRLH unit cell. C_L1=0.4pF, C_L2=0.2pF, L_L1=0.14nH, L_L2=0.25nH, C_R1=C_R2=0.52pF, L_R1=L_R2=0.075nH, R1=80Ω, R2=40Ω, G1=0.25s, G2=0.003s.*

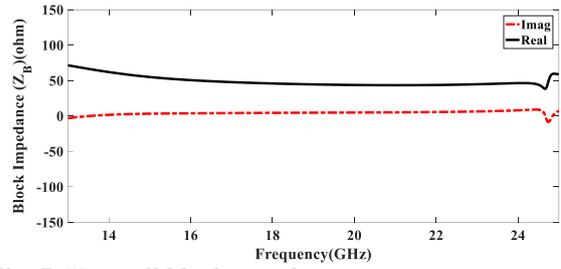

*Fig. 7. Unit cell block impedance.*

As pictured in Fig. 7, the real part of the Block impedance approximately equals 44.5Ω in the RH region, and the average of this parameter in the LH region is almost 60Ω. Based on these values for block impedance, we can design a tapered line to match the 50 Ω feeding structure with the antenna. The imaginary part of the Block impedance is nearly zero over the operational frequency range which causes a good impedance matching.

The designed LWA consists of 15 unit cells. A 50 Ω microstrip line together with a tapered line is used on each end side of the antenna for impedance matching between

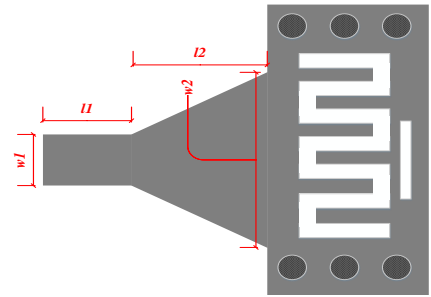

*Fig. 8. Matching structure of the proposed LWA.*

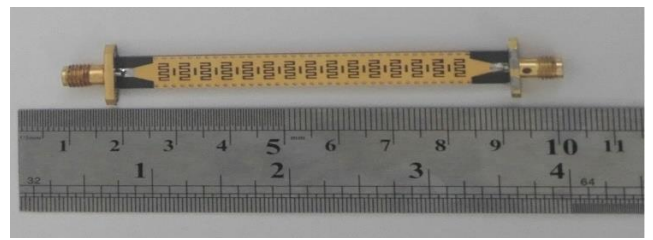

*a*

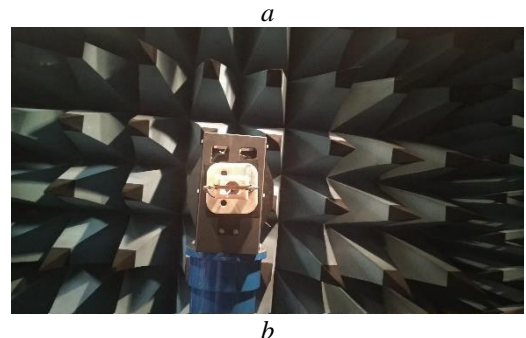

*b*

*Fig. 9. (a)* The fabricated antenna, *(b)* The antenna in the chamber

antenna and feeding structure. The dimensions of the matching structure are optimized for better impedance matching. Fig. 8 Shows the matching structure. In the antenna structure, the transverse slot adjacent to the tapered line is eliminated because of its negative effect on the antenna's function [2], [3].

## 3. Results and discussion

The final proposed antenna consists of 15 similar C_CRLH unit cells. The antenna is simulated using microwave full-wave simulation software and fabricated on the Rogers RT 5880 with a permittivity of 2.2 and a loss tangent of 0.0009. In Fig. 9, the fabricated antenna is shown with a total dimension of 72.3×8×0.8mm$^3$.

In Fig. 10, the simulated and measured scattering parameters of the antenna are plotted. According to Fig. 10, S11 extracted from the simulation is below -10 dB in the frequency range from 13.5 to 14 GHz, and is below -15 dB in the range between 14-22 GHz. Whereas in the measured $S_{11}$ graph the impedance matching frequency range is from 13 to 22 GHz. It is under -10 dB for 13-14 GHz and below -15 dB in the frequency range of 14 to 22 GHz. This difference can be originated from fabrication errors. Despite this negligible difference, there is a good agreement between simulation and measurement results.

In Fig. 11, the simulated and measured normalized radiation patterns for the backward, broadside, and forward regions are plotted. The sidelobe level is less than -10 dB and the antenna's main lobe steers versus frequency. As shown in Fig. 11, the measurement error in the right-handed region is more than the broadside and left-handed regions which might be because of higher frequency in this region and measurement equipment's sensitivity.

The antenna beam scanning angle versus frequency is plotted in Fig. 12, as shown, the simulated beam scanning angle varies from -55$^0$ to +68$^0$ (123$^0$) as the frequency varies.

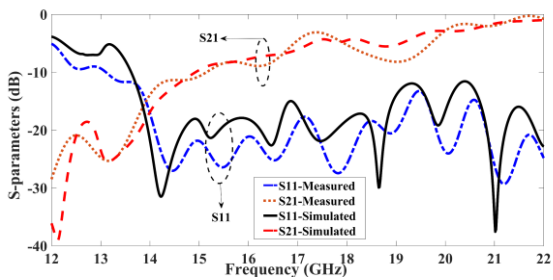

*Fig. 10. The antenna S-parameters.*

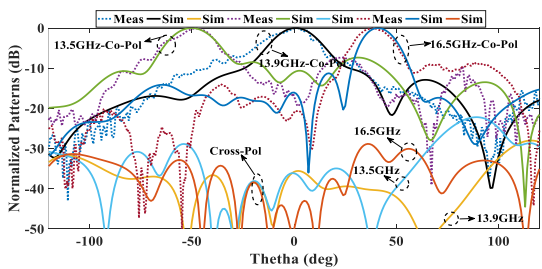

*Fig. 11. Normalized radiation pattern of the antenna.*

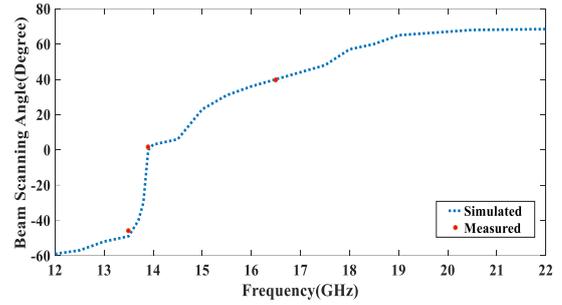

*Fig. 12. The antenna beam scanning angle.*

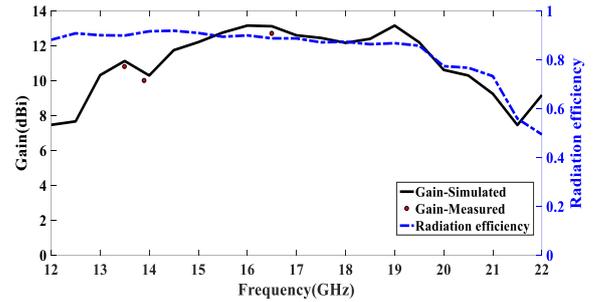

*Fig. 13. The antenna Gain (left) and Radiation efficiency (right).*

The difference between measured and simulated scanning angles is approximately 2$^0$. So, the measured beam scanning angle is from -53$^0$ to +66$^0$ (119$^0$). This wide scanning range results from wide bandwidth of the final antenna in the fast-wave region shown in Fig. 6.

Fig. 13 shows the gain and radiation efficiency of the final structure, where the simulated maximum gain and radiation efficiency are 13.16 dBi and 91.89%, respectively, and the measured maximum gain is 12.66 dBi. The measured gain maximum variation is 0.8 dBi. The average radiation efficiency over the frequency band is 84%. Besides the interdigital slot, the CRLH unit cell structure also contains transverse slots, which together improve radiation efficiency. As shown in Fig. 2(b), R2 which is the radiation resistance of a SIW structure loaded with a single transverse slot appears twice in the circuit model. Consequently, there is an increase in radiation resistance of the antenna, which increases radiation efficiency [20].

The proposed antenna's features and some of the previously developed Leaky Wave Antennas are listed in Table 2. To begin with, the maximum forward scanning angle for the proposed antenna is more than that reported in [6], [12], [13], [20], [17], [23], [24], [26], [27], [28], and [29]. Besides, the antenna's maximum backward scanning angle is -53$^0$ and is better than the backward scanning angle reported in [12], [13], [23], [24], and [25]. Altogether, the total scanning range of the antenna is improved compared to [12], [13], [17], [20], [23], [24], [25], [26], [27], [28], and [29]. The radiation efficiency of the antenna is better than that presented in [6], [13], [17], [23], [26], and [27]. Furthermore, the gain of the proposed structure is improved compared to gain in [13], [23], [24], [25], [27], and [29]. Finally, the antenna shows a broader impedance matching range compared to all of the references presented in Table 2.

**TABLE 2** Comparison with other antennas from literature

| Ref. no | Type of Antenna | Total length | BW (GHz) | $\eta_{max}$ (%) | Max Gain (dBi) | $\Delta\theta$ |
|---|---|---|---|---|---|---|
| [6] | SIW-CRLH | $4.1\lambda_0$ at 10 GHz | 8.5-13 (41.9%) | 85 | 12.8 | $\sim 130^0$ ($-70^0$ to $+60^0$) |
| [7] | Multilayer-CRLH | $4.5\lambda_0$ at 9 GHz | 8-13 (47.61%) | Not reported | 12.8 | $144^0$ ($-66^0$ to $+78^0$) |
| [12] | SIW slot antenna with CRLH TL | $2.59\lambda_0$ at 25.45 GHz | 23.95-27.725 (14.61%) | Not reported | $\sim 16.25$ | $30^0$ ($-17^0$ to $+13^0$) |
| [13] | SIW-DP CRLH | $3.43\lambda_0$ at 14.5 GHz | 10-18 (57.2%) | 80 | 11 | $82^0$ ($-42$ to $+40^0$), $4^0$ ($+33^0$ to $+37^0$) |
| [17] | HWSIW CRLH LWA with Spiral Resonator | $4.85\lambda_0$ at 4.3 GHz | 13.5-17.8 (27.5%) | 85 | 16 | $86^0$ ($-66^0$ to $+20^0$) |
| [20] | EMSIW CRLH | $5\lambda_0$ at 11.25 GHz | 9-13.5 (40%) | 96 | 17.96 | $107^0$ ($-64^0$ to $+43^0$) |
| [23] | Asymmetric SIW LWA | $\sim 10.3\lambda_0$ at 14 GHz | 12-16 (28.57%) | 85 | 12.5 | $59^0$ ($-32^0$ to $+27^0$) |
| [24] | 1-D Slot Array LWA | $9.56\lambda_0$ at 11.4 GHz | 8.5-14.3 (50.87%) | Not reported | 12 | $87^0$ ($-51^0$ to $+36^0$) |
| [25] | SIW-CRLH with H shape unit-cell | $3.73\lambda_0$ at 9.1 GHz | 7.6-10.6 (32.97%) | 94.98 | 10.5 | $109^0$ ($-38^0$ to $+71^0$) |
| [26] | SIW with tapered half-wavelength transmission line | Not reported | 11.2-15.6 (32.83%) | >80 | 15 | $112^0$ ($-55^0$ to $+57$) |
| [27] | Non-uniform CRLH HMSIW | $3.07\lambda_0$ at 8 GHz | 7-10.2 (40%) | 85 | 9.8 | $117^0$ ($-66^0$ to $+51^0$) |
| [28] | S-L SIW | $12.37\lambda_0$ at 14 GHz | 10.5-17.5 (50%) | Not reported | 16.5 | $84^0$ ($-64^0$ to $+20^0$) |
| [29] | SIW with Eye-Shaped Transverse slot | $4.84\lambda_0$ at 11 GHz | 9.1-13 (35.3%) | >90* 65 (mea) | 11.7 | $95^0$ ($-56^0$ to $+39^0$) |
| **This work** | **SIW-C CRLH** | **3.34 $\lambda_0$ at 13.9 GHz** | **13-22 (51.42%)** | **91.89** | **12.66** | **$\sim 119^0$ ($-53^0$ to $+66^0$)** |

DP: DOUBLE PERIODIC, C-CRLH: COMPACT CRLH, EM: EIGHT-MODE, S-L: SINGLE LAYER, * WITHOUT SMA CONNECTOR LOSS

## 4. Conclusion

A wide-angle SIW Leaky Wave Antenna with improved gain, radiation efficiency, and bandwidth based on a C-CRLH unit cell has been developed in this study. The unit cell contains one interdigital slot and two transverse slots. The gain, radiation efficiency, and continuous beam steering capability of the developed antenna are examined. Further, in the unit cell, the relationship between the distance between the interdigital slot and transverse slots with coupled power to the free space is examined to determine the general length of the LWA with more accuracy. The designed antenna provides beam scanning from $-53^0$ to $+66^0$ across the impedance bandwidth of 13 to 22 GHz with a maximum gain of 12.66 dBi. Therefore, it is suited for wideband applications, such as imaging, due to its wideband impedance matching and wide beam scanning range.